\documentclass[aip,psfig,twocolumn,showpacs,superscriptaddress,reprint,floatfix]{revtex4-1}

\usepackage{amsmath}
\usepackage{mathtools}
\usepackage{amssymb}
\usepackage{color}
\usepackage{graphics}
\usepackage{epsfig}
\usepackage[normalem]{ulem}                     % to strike the words
\usepackage{cancel}
\usepackage[mathlines]{lineno}
\graphicspath{{Images_dusty_plasma/}}
\usepackage{xtab,afterpage,longtable}
\usepackage{booktabs}   
\usepackage{ltablex}      
\relax
\usepackage{float}
\begin{document}
%\maketitle
%%%%%%%%%%%%%%%%%%%%%%%%%%%%%%%%%%%%%%%%%%%%%%%%%%%%%%%%%%%%%%%%%%%%%%%%%%%%%%%%%%%%%%%%%%%%%%%%%%%%
\preprint{AIP/123-QED}
\title{Observation of Kolmogorov turbulence due to multiscale vortices in dusty plasma experiments}

\author{Sachin Sharma}
 \affiliation{Indian Institute of Technology Jammu, Department of Physics, Jammu, 181221, India}
\author{Rauoof Wani}
\email{abrauoofwani@gmail.com}
\affiliation{Indian Institute of Technology Jammu, Department of Physics, Jammu, 181221, India}
\author{Prabhakar Srivastav}
\affiliation{Institute for Plasma Research, Bhat, Gandhinagar, Gujarat, 382428, India}
\author{Meenakshee Sharma}
\affiliation{Bellatrix Aerospace Pvt. Ltd., \#22, 5th floor, Sankey road, Bangalore, 560020, India}
\author{Sayak  Bose}
\affiliation{Princeton Plasma Physics Laboratory, Princeton, NJ 08540, USA.}
%\affiliation{Columbia Astrophysics Laboratory, Columbia University, 550 West 120th Street, New York, NY 10027, USA}
\author{Yogesh Saxena}
\affiliation{Institute for Plasma Research, Bhat, Gandhinagar, Gujarat, 382428, India}
\author{Sanat Tiwari}
\affiliation{Indian Institute of Technology Jammu, Department of Physics, Jammu, 181221, India}
\date{\today}
% It is always \today, today,
%  but any date may be explicitly specified
%%%%%%%%%%%%%%%%%%%%%%%%%%%%%%%%%%%%%%%%%%%%%%%%%%%%%%%%%%%%%%%%%%%%%%
\begin{abstract}
We report the experimental observation of fully developed Kolmogorov turbulence originating from self-excited vortex flows in a three-dimensional (3D) dust cloud. The characteristic -5/3 scaling of 3D Kolmogorov turbulence is consistent in both the spatial and temporal energy spectra within a statistical variation of experimental data. Additionally, the 2/3 scaling in the second-order structure function further supports the presence of Kolmogorov turbulence. We also identified a slight deviation in the tails of the probability distribution functions for velocity gradients, a reflection of intermittency. % The dust cloud is formed in the diffused region away from the electrode. 
The experiment showed the formation of a dust cloud in the diffused plasma region away from the electrodes.
%The cloud in levitated above the glass surface and is away from the electrode.
The dust rotation was observed in multiple experimental campaigns under different discharge conditions at different spatial locations and background plasma environments.
\end{abstract}

\maketitle
\section{Introduction}
%~~~~~~~~~~~~~~~~~~~~~~~~~~~~~~~~
\paragraph*{}
Laboratory dusty (complex) plasmas comprise of  electrons, ions, neutral gas particles, and highly charged dust particles~\citep{melzer2019physics,PK_shukla_2015}.  Dust particles generally have sizes up to tens of $\mu$m and typically attain a high negative charge of $10^4$e due to the high mobility of electrons. However, under atypical conditions, the dust particles can also acquire positive charges transiently or stably, as reported in various afterglow dusty plasma experiments~\citep{Neerajchaubeypositivechargingafterglow,Samarian_etal}. Their dynamics are slow (a few tens of Hz) compared to electron and ion timescales. The length scales  usually  ranges from a few millimeters to a few centimeters~\citep{melzer2019physics}. The  slow dynamics and convenient length scales render it an ideal experimental platform for accurately monitoring complex physics phenomena. Some examples are phase transition, longitudinal and transverse waves, solitons, shocks, vortices, and complex flows~\citep{Bailung_pop_2020,Bandyopadhyay_PRL_2008,Jaiswal_PSST_2016,Mangilal_DAW,pramanik_PRL_2002,Heinrich_PRL_2009,Garima_PRE_2021,manjit_pop_15,Bose_2019}.

Dusty plasma experiments with precision imaging diagnostics enable the study of complex flows. It includes two and three-dimensional laminar, sheared, rotating, chaotic, and turbulent flows~\citep{Nosenko_PRL_2004,pramanik_PRL_2002, Jaiswal_PSST_2016,GAVRIKOV_PRA_2005,Mangilal_POP_2020}. These low Reynolds number flows can be classified as either hydrodynamic or visco-elastic. This classification depends on the interaction strength of the charged dust particles relative to their thermal energy in the plasma.
Dusty plasma flows are often observed in experiments due to the interplay of various forces acting on dust particles.
These flows can be tracked from particle scales to hydrodynamic scales~\cite {Feng_RSI_2007, Thomas_IEEE_2010, Boesse_ASR_2004}.
In this study, we observed the turbulent flow caused by the dust vortex, particularly within the diffuse region of a DC glow discharge plasma. 
We found the alignment with Kolmogorov power law scaling in both spatial and temporal domains of the energy spectrum, thus demonstrating Kolmogorov turbulence in afterglow-like dusty plasma conditions.

Dusty plasma experiments provide a unique platform for studying turbulent flows across both macroscopic (system size) and microscopic (average interparticle separation) scales. This precision is achieved by tracking individual dust particles' trajectories, even in the liquid phase. These experiments serve as an alternative to Navier-Stokes equation-based numerical models, which depend on the continuum approximation. Such models are computationally expensive because they require numerous grid points to resolve scales close to the particle level~\citep{MARION_1998503,wani2023thermalization}.

Another notable aspect of dusty plasma experiments is that they can easily be formed in a strongly coupled regime. Turbulent mixing within such a regime holds significance in astrophysical flows, inertial confinement fusion, and other areas~\citep{das2010turbulence,weber2014inhibition,ongena2016magnetic,hardman2019multiscale,conway2008turbulence}. Furthermore, the strong coupling phenomenon is recognized for introducing viscoelastic features in dust flows~\citep{V_dharodi_PoP_2024}. Consequently, these experiments serve as a foundation for investigating turbulence within the low Reynolds number regime, often called viscoelastic turbulence \citep{Datta_PRF_2022}.
Finally, dusty plasma experiments provide a test bed to observe the effect of gravity on turbulent mixing. Theoretically, gravity is predicted to steepen energy transfer scaling in at least two-dimensional flows~\citep{Bolgiano1959}. In our experiment, the dust flows, and rotation forms naturally at a particular set of discharge conditions.  We will examine the observed flow for turbulence characteristics based on the possibilities mentioned earlier. 
\paragraph*{} 
Nonlinear mixing and turbulence in dusty plasma
can broadly have two origins: nonlinear wave interaction and vortex formation~\citep{schwabe_turbulence_dusty_plasma,schwabe_IEEE,tiwari2015turbulence, Suraj_kalita, E_joshi_PRR, Bailung_pop_2020,bajaj_PRE_turbulence,kostadinova_PoP_2021}. In numerical modeling, turbulent mixing in dust fluid occurs by initiating energy/perturbation in vortices of system size or by introducing various seed instabilities~\citep{RAUOOF_2022_Nature}. 
The main question is whether the turbulence characteristics in strongly coupled plasmas resemble or differ from those in hydrodynamic fluids. In the context of turbulence in hydrodynamic fluids, Kolmogorov proposed that the energy spectrum $E(k)$ follows a $k^{-5/3}$ scaling in the inertial range. It applies to any homogeneous, isotropic turbulent flow forced at a large scale. Here, $k$ is the wavenumber associated with the particular eddy in the turbulent fluid system, and $E(k)$ is the energy associated with the respective eddy~\cite{kraichnan1974kolmogorov}. 

\paragraph*{}
For dusty plasmas, Schwabe \textit{et al.}\cite{schwabe_turbulence_dusty_plasma} experimentally demonstrated a dual cascade in the spectrum. They found that $E(k) \propto k^{-5/3}$ for $k < k_{l}$ and $E(k) \propto k^{-3}$ for $k > k_{l}$, where $k_{l}$ is an intermediate forces wavenumber. Their results are consistent with theoretical predictions of two-dimensional forced turbulence~\citep{kraichnan_1973,kraichnan_1967_POF,kraichnan1974kolmogorov}. Tsai and Lin I~\citep{Tsai_PRE_2014} demonstrated that the dust acoustic wave turbulence exhibits $f^{-2.6}$ scaling in a frequency spectrum. Using kinetic numerical simulations, Joshi \textit{et al.}~\citep{E_joshi_PRR} demonstrated turbulent mixing due to the vortex motion in the dust flow past the obstacle. Their results follow the hydrodynamic Kolmogorov turbulence scaling of $-5/3$ in their energy spectrum in both the time and spatial domains. Using molecular dynamics simulation,~\citet{Rauoof_PoP_2024} demonstrated the turbulence in a dusty plasma system using Rayleigh-Taylor instability as an initial condition. Their spectra align with the  Bolgiano-Obukho theory, a typical two-dimensional Rayleigh-Taylor turbulence.
While all previous experimental works report wave turbulence, this work is the first to demonstrate fully developed Kolmogorov turbulence in dusty plasma due to multi-scale vortex flow. The vortex flow appears stationary at the largest scale of energy infusion. We could show the scaling of the energy cascade close to $-5/3$ in the wavevector and frequency domains.
\paragraph*{}
In this paper, we report a few large-scale vortices and small-scale eddies in a dust cloud. It levitates in the diffused plasma far from the electrode in a DC discharge. The dust cloud remains stable for several hours due to the balance of electrostatic $(q\vec{E})$ and gravitational $(m\vec{g})$ forces. This stability is maintained as long as an external source continuously provides energy, and the dust cloud resides in an energy bath of background plasma. The dust cloud is 3D in shape. However, the image analysis and Particle Image Velocimetry (PIV) are carried out on the basis of the two-dimensional imaging of the cloud. We estimate two-order-of-magnitude background plasma density variations between the electrode and the diffused plasma region. The observed vortex is reproducible on the SPD-I device~\citep{sharma2023shivalik}. We presented the turbulence features in the rotating dust cloud of the diffused plasma region. Understanding the flow dynamics in diffused plasma regions is interesting due to their relevance to the interstellar universe and the lunar surface, which have 
minor density variations. We observe a scaling consistent with Kolmogorov's ${-5/3}$ in the frequency and spatial domains of the energy spectra. Also, in all energy spectra calculations, the kinetic energy is computed using only the fluid velocity components without incorporating the masses of fluid elements. To further support our results, we have demonstrated the second-order structure functions following the Kolmogorov law and the Probability distribution function (PDF) of velocity gradients.  
\paragraph*{}
The paper is organized as follows: section \ref{exp_device} details of the device, image processing technique, and estimated plasma parameters. Section \ref{res_dis} provides observations and analysis followed by an associated discussion. Finally, section \ref{conclusion} summarizes the outcomes with future possibilities. 
%~~~~~~~~~~~~~~~~~~~~~~~~~~~~~~~
%~~~~~~~~~~~~~~~~~~~~~~~~~~~~~~~
\section{Experimental details}
\label{exp_device}
\paragraph*{}
%%%%%%%%%%%%%%%%%%%%%%%%%%%%%
\begin{figure}
\includegraphics[width =\columnwidth]{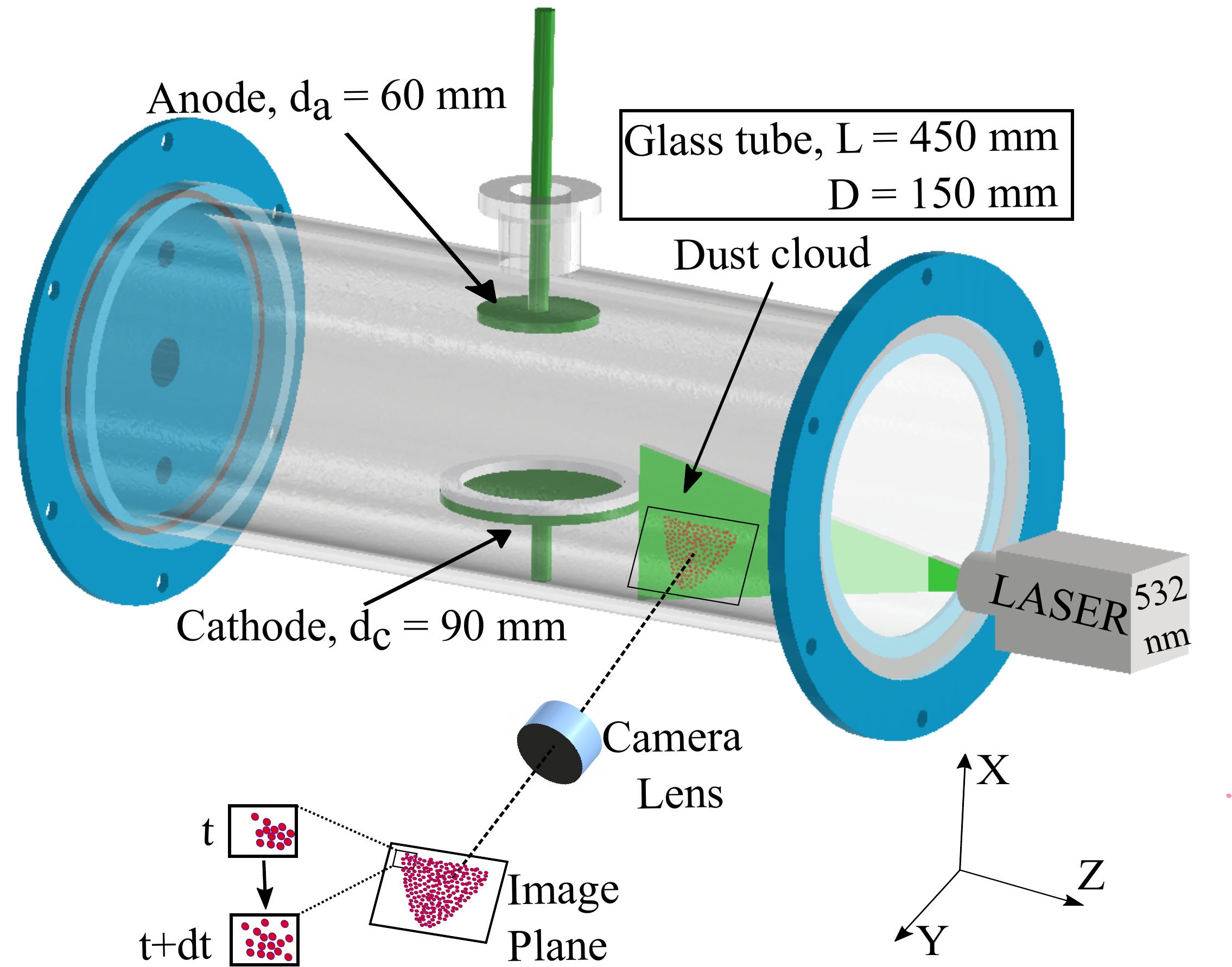}
\caption{The schematic of SPD-I shows a laser sheet in the X-Z plane and a camera lens positioned perpendicularly to capture images. These images are used for PIV analysis to determine velocity fields in the X and Z directions. For detailed device dimensions, see~\citet{sharma2023shivalik}.}
\label{Fig1}
\end{figure}
%%%%%%%%%%%%%%%%%%%%%%%%%%%%%
The experiments are performed on a newly commissioned laboratory plasma device, Shivalik Plasma Device-I (SPD-I)~\cite{sharma2023shivalik}. This plasma chamber is a horizontally laid (Z-axis) glass cylinder with two asymmetric electrodes kept vertically in a parallel plate arrangement, as shown in Fig.~\ref{Fig1}. The glass chamber is electrically floating using Teflon connectors, with the top electrode as an anode and the bottom grounded electrode as a cathode. In the present experiments, we separated both electrodes by 7.5 cm. Figure~\ref{Fig1} depicts this DC discharge-based experimental set-up. The details on configuration, including the device dimensions, electrodes, and diagnostics, are in~\citet{sharma2023shivalik}. We introduced Argon, the working gas, into the vacuum chamber through a port adjacent to the pumping port to prevent any directed flow of neutral gas into the experimental region of interest. This approach ensures that only diffused gas enters the system~\citep{Kaur_Pramana_2016}.
We repeatedly purged the device with Argon gas at high pressures ($\sim5$~mbar) for at least 10 minutes before maintaining working pressure to conduct the experiment. The plasma was formed in an Argon gas background at a low working pressure of 0.12 mbar.

Polydispersive kaolin particles, ranging in size from 4 to 8 $\mu$m, are dispersed over the cathode before creating the vacuum. We performed a slow initial pumping to ensure the Kaolin particles remained inside the device. As soon as plasma creation takes place, the dust particles covering the cathode, along with some of its floating fraction, become charged. A void quickly forms over the cathode because of the competition between the ion drag force and the radial electric field. During the void formation, the electrode expels almost all floating dust particles. However, these dust particles then form a 3D cloud in the diffused plasma region over the extended sheath on the glass surface. The 3D cloud forms, possibly for two reasons: first, because of polydispersive Kaolin particles, and second, because of the 3D potential well formation in the diffused plasma region. The second reason is the predominant cause of the 3D dust cloud in the present experiment. 
We validated it using monodispersive MF particles in the SPD-I device~\cite{sharma2023shivalik} and a similar 3D dust cloud formed.
Three forces may contribute to this potential well formation. 1) A radially outward force on dust due to void formation over the cathode. 2) The force generated by the lower glass surface's sheath electric field. 3) The force emanating from the electric field profiles of the end-flange sheath. 
\paragraph*{}
A green LASER with a 532 nm wavelength and a variable power capacity of up to 300 mW illuminates the dust particles. A cylindrical lens is placed before the LASER to convert it into a thin sheet of thickness 150 $\mu$m. A high-speed camera captures the illuminated dust particles at a 90-degree angle to the LASER sheet (X-Z plane). We have also made a few observations by keeping the LASER sheet in the Y-Z plane to support the three-dimensional nature of the dynamics. We have used two different cameras. The PCO Edge 5.5 is utilized for recording up to 100 fps with a resolution of 2560 $\times$ 2180 pixels, chosen for its superior spatial resolution. The Phantom v2512 records up to 1000 fps with a resolution of 1064 $\times$ 1064 pixels, selected to achieve a superior temporal data profile. PIV has served as the primary tool for extracting velocity fields from the recorded image sequences (detailed in subsection ~\ref{IPP}). We also extract particle positions and velocities by tracking particles over time for at least two experimental datasets. We used these particle data sets to validate PIV results and predict the static structural properties of dust clouds.
%%%%%%%%%%%
%%%%%%%%%%%%%%%
\begin{figure}
\includegraphics[width =0.9\columnwidth]{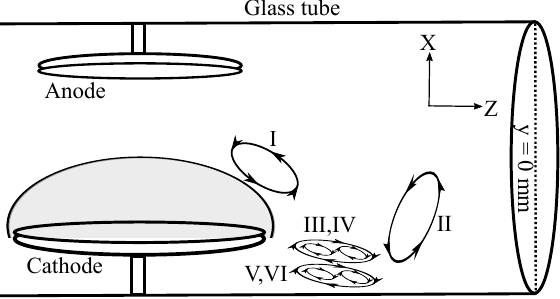}
\caption{Schematic of right-half of the plasma chamber showing the tentative locations of dust cloud rotations from different campaigns as per Table~\ref{table1_sets}.
}
\label{Dust_cloud_location}
\end{figure}

%%%%%%%%%
\subsection{Datasets and plasma parameters}
\paragraph*{}
We conducted multiple campaigns to study flows and turbulence to ensure the reproducibility and consistency of observed dust rotation phenomena. A total of six independent experimental campaigns were conducted at the same working pressure of 0.12 mbar, as listed in Table~\ref{table1_sets}. All of these campaigns have been performed using polydispersive Kaolin microparticles. The schematic in Fig.~\ref{Dust_cloud_location} shows tentative dust rotation locations from different campaigns. These campaigns successfully captured the X-Z plane of the dust cloud, ensuring that the laser sheet remained at the center of the glass flange at location $y=0$. The hemispherical, shaded region represents the dust void. In each campaign, we observed dust rotation with variations in flow characteristics. Specifically, I, II, III, and VI form in reasonably different plasma environments. We anticipate an interplay of different sets of forces~\cite{Kaur_POP_2015a,Chai_POP_2016, Vaulina_NJP_2003,Mangilal_POP_2018,Samarian_Scripta_2002} to be responsible for individual vortices in the extended diffused region. It is interesting to note that in all these conditions, the nature of turbulence almost follows the standard forced Kolmogorov theory. 
%%%%%%%%%%%%%%%%%%%%%%%%%%
%%%%%%%
\begin{table}[htbp]
\centering
\caption{Campaign details: The table provides the discharge voltage, $V_d$, and the tentative cloud span, $S_{xz}$. It also includes their horizontal distance from the cathode, $d_{hc}$, and the vertical distance from the glass surface, $d_{vg}$.}
\vspace{0.2 cm}
\label{table1_sets}
\begin{tabular}{c|cccc}
C No. & $V_d$(Volts) & $S_{xz}$($\mathrm{mm^2}$) & $d_{hc}$(mm) & $d_{vg}$(mm) %\vspace{1.5mm}
\\ \hline
I & 270  & $14 \times 26$ & 40   & 56 \\
II & 360  & $45 \times 44$ & 73 & 1 \\
III & 565  & $7\times 55$ & 55 & 1.1 \\
IV & 536 & $12 \times 35$ & 50 & 1 \\
V & 538  & $11\times 35$ & 60 & 1   \\
VI & 570  & $10\times 35$ & 60 & 0.9 \\
\end{tabular}
\end{table}

%
%
%%%%%%%%%%%%%%%%%%%%%%%%%%%%%
\begin{table}[htbp]
  \centering
 \caption{Relevant plasma parameters for the present study.}
 \vspace{0.2cm}
  \label{table2_plasma}
  \begin{tabular}{|p{3.5cm}|p{2cm}|p{2cm}|}
    \hline
    \textbf{Parameters}  & \textbf{Main plasma} & \textbf{Diffused plasma} \\
    \hline 
    Electron temperature, $T_e$  & $1.5-4$ eV & $3-7$ eV\\
    \hline
    Plasma density, $n_{e,i}$  & [1 - 3] $\times$ 10$^{9}$ cm$^{-3}$ & [4 - 8] $\times$ 10$^{7}$ cm$^{-3}$\\
    \hline 
    Debye length, $\lambda_D$  & $23-40$ $\mu$m & $143-200$ $\mu$m\\
    \hline 
    Ion plasma frequency, $\nu_i$  &  $5.9 \times 10^{7}$ Hz  &  $5.9 \times 10^{6}$ Hz \\
    \hline
    Electron plasma frequency, $\nu_e$  & $4 \times 10^{8}$ Hz & $4 \times 10^{7}$ Hz\\
   \hline
  \multicolumn{2}{|c|}{Dust particle mass density, $\rho_m$}  &  $2.5$ g cm$^{-3}$\\
    \hline
   \multicolumn{2}{|c|}{Dust size (radius), $r_d$}  & $2.5-4$ $\mu$m \\
    \hline 
    \multicolumn{2}{|c|}{Dust mass, $M_d$} & $[1.6 - 6.7] \times 10^{-13}$ $\mathrm{kg}$ \\
    \hline 
    \multicolumn{2}{|c|}{Dust average interparticle separation, $a$}  & %$224 \pm xx$ $\mu$m $^\#$ \\
    $220-300$ $\mu$m \\
    \hline 
    \multicolumn{2}{|c|}{Dust density  in plasma, $n_d$} & 
    %$2.314 \pm xx \times 10^4$ /cm$^{-3}$ $^\#$ \\
    $[0.9-2.24] \times 10^4$ cm$^{-3}$ \\
    \hline 
    \multicolumn{2}{|c|}{Shielding parameter, $\kappa = a/\lambda_D$}  & $1.2-2.2$ \\
    \hline 
    \multicolumn{2}{|c|}{Typical dust charge, $Q_d = 1400 \times (2r_d) T_e$ }  & $[2.1-5.6] \times 10^4$ e \\
    \hline 
    \multicolumn{2}{|c|}{Dust plasma frequency, $\omega_{pd}$ } & 55-67 Hz \\
    \hline
    \multicolumn{2}{|c|}{Dust temperature, $T_d$}  & $0.6 - 2.2$ eV  \\
    \hline 
    \multicolumn{2}{|c|}{Coupling parameter, $\Gamma$ } & $1500-4000$\\
    \hline 
    \multicolumn{2}{|c|}{Epstein damping rate~\citep{Epstein_1924,Pustylnik_PRE_2006}, $\gamma_{ep}$ } & $8~\mathrm{s^{-1}}$\\
    \hline 
    \end{tabular}
\end{table}
%%%%%%%%%%%%%%%%%%%%%%%%%%%%%%%%%%%%%%%%%
\begin{figure}
\includegraphics[width = 0.9\columnwidth]{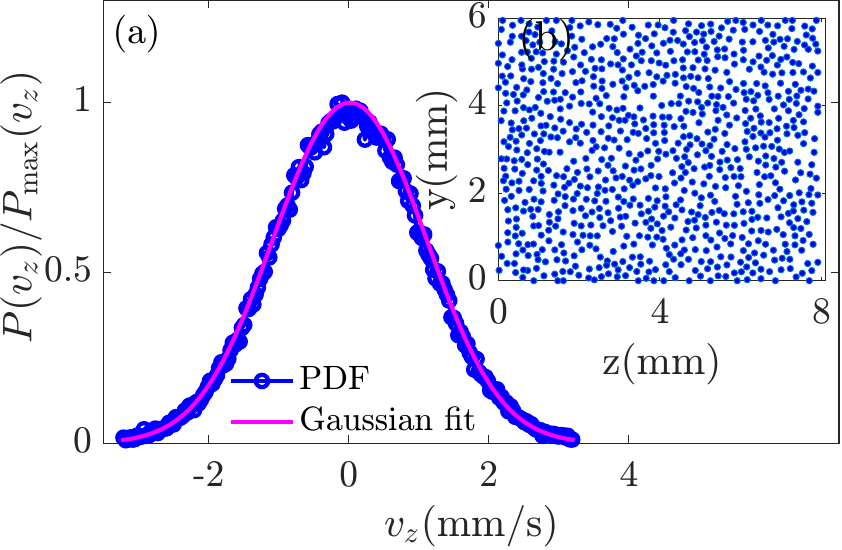}
\caption{(a) Normalized probability distribution function of particle velocities fitted with the Gaussian distribution. The charged dust particle temperature is obtained from full-width half maxima of fitted Gaussian. 
(b) Particle positions in the YZ plane from campaign V%, with data extracted using ImageJ software~\cite{Fiji_imageJ}
. The extraction region corresponds to a visually no-flow equilibrium zone, capturing dominantly the thermal motion.}
\label{temperature_dust}
\end{figure}
%%%%%%%%%%%%%%%%%%%%%%%%%%%%%%%%%%%%%%%%%%%%%%
%%%%%%%%%%%%%%%%%%%%%%%%%%%%%%%%%%%%%%%%%%%%%%%
\begin{figure}
\includegraphics[width =0.8\columnwidth]{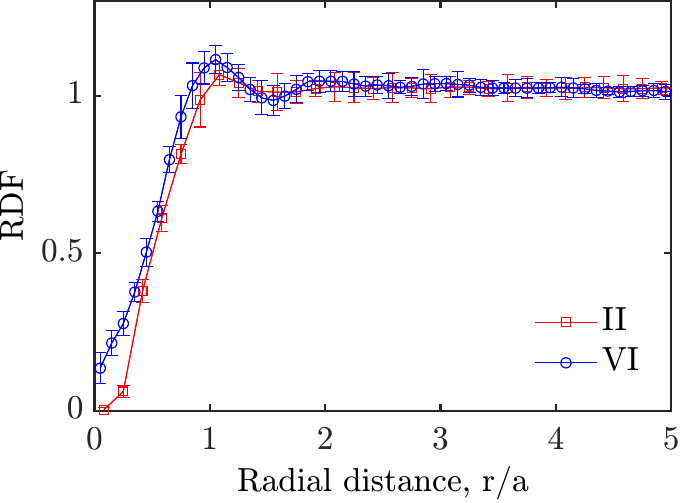}
\caption{The radial distribution functions, $g(r)$ obtained from the particle data for two datasets.}
% For comparison, we also plotted the RDF from a molecular dynamics simulation (red dash-dotted line) with parameters $\Gamma = 26$ and $\kappa = 5.0$
\label{RDF}
\end{figure}
%%%%%%%%%%%%%%%%%%%%%%%%%%%%%%%%%%%%%%%%%%%%%
\paragraph*{}
Table~\ref{table2_plasma} enlists parameters for the background plasmas and the dust cloud. The electron temperature, $T_e$, and the plasma density, $n_{e, i}$ (in quasineutral plasma), were measured using the single Langmuir probe above the electrode and in the diffused plasma region in the absence of dust~\cite{Bose_2017}. Over the cathode, $T_e$ ranges $1.5-4$ eV from the center to the edge. In the diffused region where dusty plasma typically forms, $T_e$ is relatively high, ranging from 3 - 7 eV at different locations within the domain span. 
The plasma density varies by up to two orders of magnitude from the electrode to the diffused region of interest, as shown in the table.  
%%%%%%
\paragraph*{}
We used a microscope to obtain the kaolin particles' size distribution. Most particles ranged from 5 to 8 $\mu$m, although there were a few outliers with sizes as large as 50 $\mu$m. Based on the known mass density of kaolin used in the experiments, the typical mass of dust particles ranged from $1.6 \times 10^{-13}$ to $6.7 \times 10^{-13}$ kg. 
Out of the six campaigns, we could track particle positions with reasonable accuracy using the ImageJ software~\cite{Fiji_imageJ} for two campaigns.
We used the particle data from Fig.~\ref{temperature_dust}(b) for 145 successive frames and a particle tracking algorithm to gather particle velocities (in mm/s).
Determining the accurate average interparticle separation, $(a)$, for dust clouds composed of particles of varying sizes and experiencing different spatial plasma environments is challenging.
However, we calculated ``$a$" from regions near vortex motion with visually no flow but only thermal motion. We obtained an estimated value of $a \approx 224~\mu$m and a typical dust density, $n_d \approx 2.3 \times 10^4$ cm$^{-3}$ in the diffused plasma region. Figure~\ref{temperature_dust}(b) depicts one such particle arrangement (from Campaign V) we used to estimate $a$.
%%%%%
\paragraph*{}
Dust particles typically have a charge between [2 - 5] $\times 10^4$$e$. The floating potential on a spherical dust particle is the basis for the collisionless orbital motion limited model, which is used to compute the charge. 
We fitted the theoretical Gaussian velocity distribution to the Probability Distribution Function (PDF) of the particle velocity data (averaged PDF from 145 frames), which we plotted in Fig.~\ref{temperature_dust}(a). A theoretical fitting in comparison with experimental data yields dust temperature ranges of 0.6 - 2.2 eV. We see that in a three-dimensional situation, the dust temperature will vary since it is computed using a two-dimensional velocity field. 
For two datasets (II and VI), we additionally computed the 2D radial distribution function (RDF). The dynamics are thermal, and there is no bulk flow in the area used for this computation. With a tiny initial peak, the RDF in Fig.~\ref{RDF} illustrates the equilibrium dust dynamics in a somewhat liquid regime.
\subsection{Image processing and PIV}
\label{IPP}
\paragraph*{}
Image processing and PIV are crucial for accurately extracting fluid velocity in the dust cloud. We detail the processes adopted from recorded videos to ascertain velocities over a square grid. We processed video data from six campaigns (I to VI) to study fluid velocities and turbulent mixing. Campaigns I and II were recorded at 25 fps, III and IV at 101 fps, and V and VI at 125 fps.
All original images were pre-processed using the open-source ImageJ software by converting them to grayscale. Thresholding was then applied to convert all pixels with an intensity range of 50-220 to maximum brightness, while other pixel intensities were set to zero. A convolution filter sharpened the particles for better clarity, especially given the significant particle-size asymmetry. This data was then ready to export for PIV studies.
%%%%%%%
\paragraph*{}
We used MATLAB-based PIVlab software for the PIV analysis of the pre-processed image data~\cite{Thielicke-2021}. After importing the images, we selected the square region of interest (RoI) to study vortex and turbulence flows. Unwanted regions (blank spaces with no dust) were removed by masking.
 For the PIV analysis, we used a $32 \times 32$ pixel interrogation area with a 75\% overlap of windows. The FFT-based cross-correlation algorithm calculated velocities (pixels/frame) for all windows within the RoI. The velocity data was then calibrated based on actual experimental dimensions, using pre-marked references in the images for length and fps for timescales. Unwanted high-velocity data significantly higher than the bulk flow speed were removed using velocity histogram cropping. Finally, we obtained N-1 sets of velocity field data from N images and used them to profile velocity fields, vorticity, and energy spectra.
 %%%%%%%%%%%%%%%%
\begin{figure}
\includegraphics[width =  0.95\columnwidth]{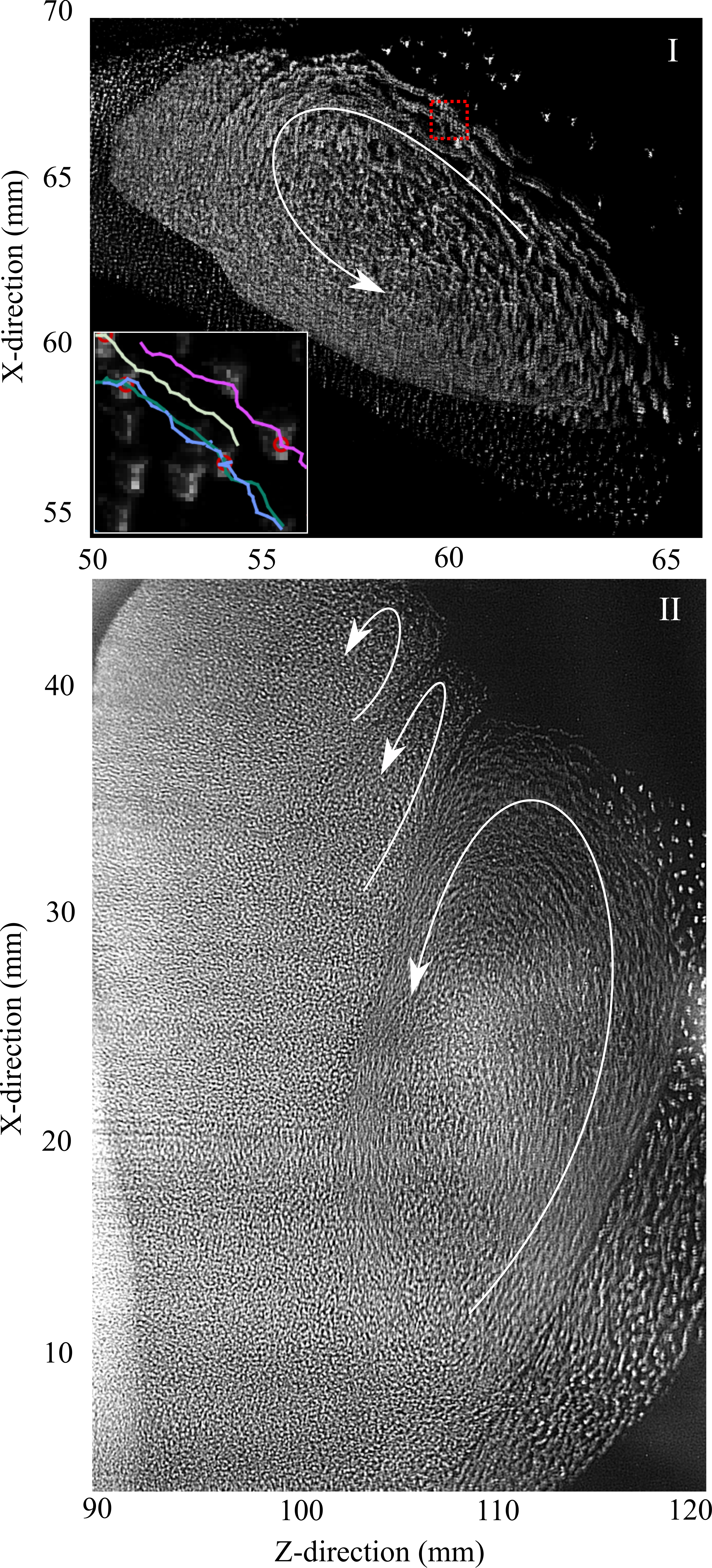}
\caption{Time-elapsed images (20 consecutive frames for each case) of dust flow in campaigns I and II. Their locations and discharge parameters are as per Table~
\ref{table1_sets}. The inset for I highlights the trajectories of a few selected particles within the cloud.}
\label{Fig_2}
\end{figure}
%%%%%%%%%%%%%%
%\end{document}
\section{Results and Discussions}
\label{res_dis}
%%%
\paragraph*{}
We report the observation of self-excited dust rotation dynamics in the 3D dust cloud within a diffused plasma. We analyze its properties to understand its turbulence characteristics.
The following is a detailed discussion of the results.
\begin{figure}
\includegraphics[width =\columnwidth]{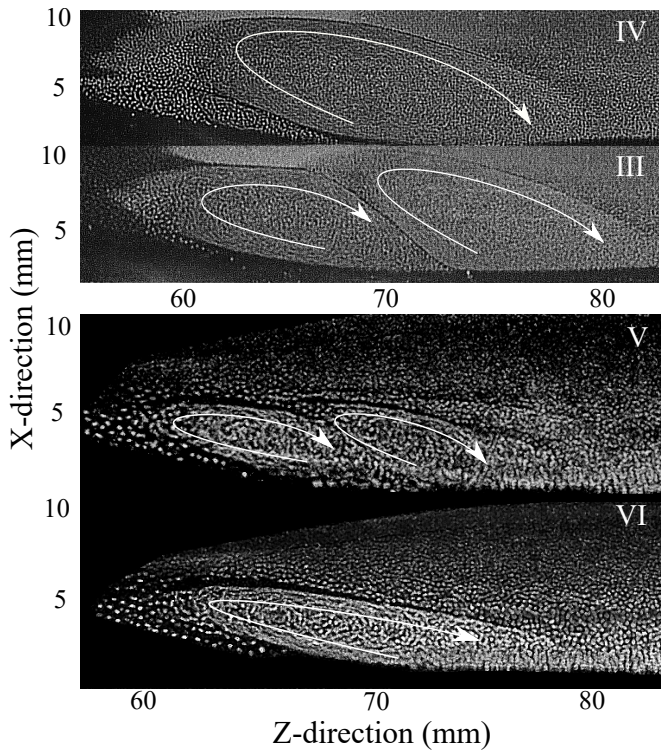}
\caption{Time-elapsed images (20 consecutive frames for each case) of dust flow in campaigns III-VI. Their locations and discharge parameters are as per Table~
\ref{table1_sets}.}
\label{Fig_222}
\end{figure}
%%%%%%%%%%%%%
\subsection{Dust vortex formation}
\label{dust_rot}
\paragraph*{}
%%%%%%%%%%%%%%%%%%%%%%%%%%%
\begin{figure}
\includegraphics[width = 
\columnwidth]{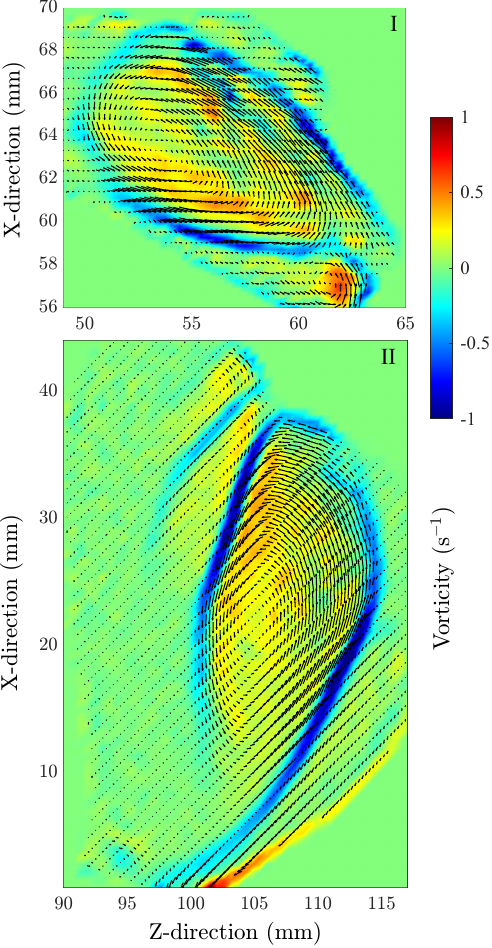}
\caption{Vorticity profile for dust flow in case of campaigns I and II. The dust cloud formed above the glass surface at low discharge voltages.}
\label{Fig_4a}
\end{figure}
%%%%%%%%%%%%%%%%%%%%%%%%%%%
%%%%%%%%%%%%%%%%%%%%%%%
 \begin{figure*}
 \includegraphics[width = 
 \textwidth]{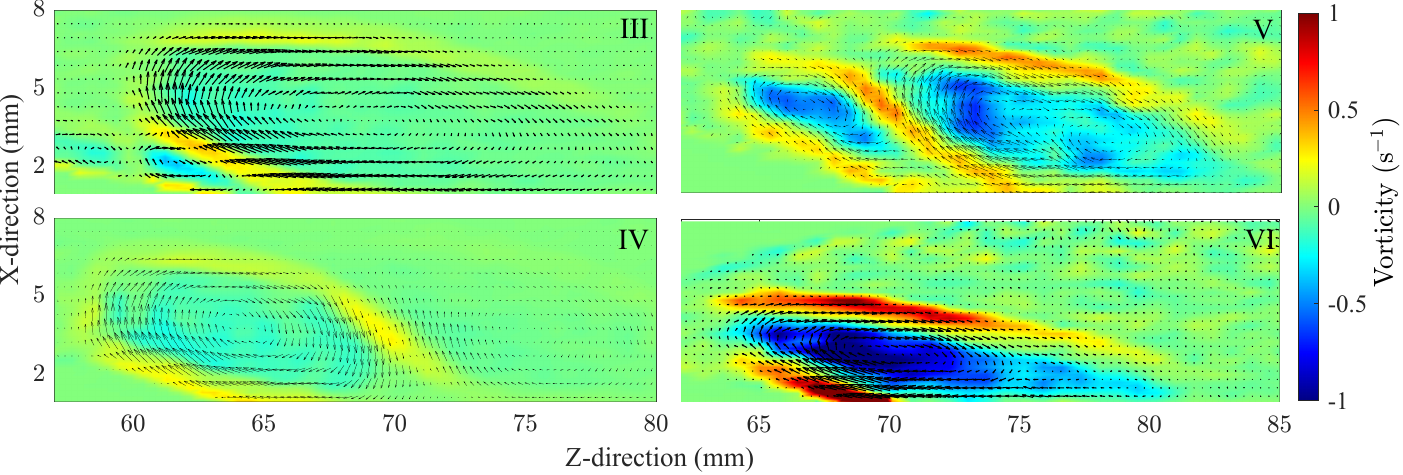}
 \caption{Vorticity profile for dust flow in regions III-VI of the dust cloud formed above the glass surface at high discharge voltages.}
 \label{Fig_4}
 \end{figure*}

 \begin{figure*}[btp]
\includegraphics[width =0.9\textwidth]{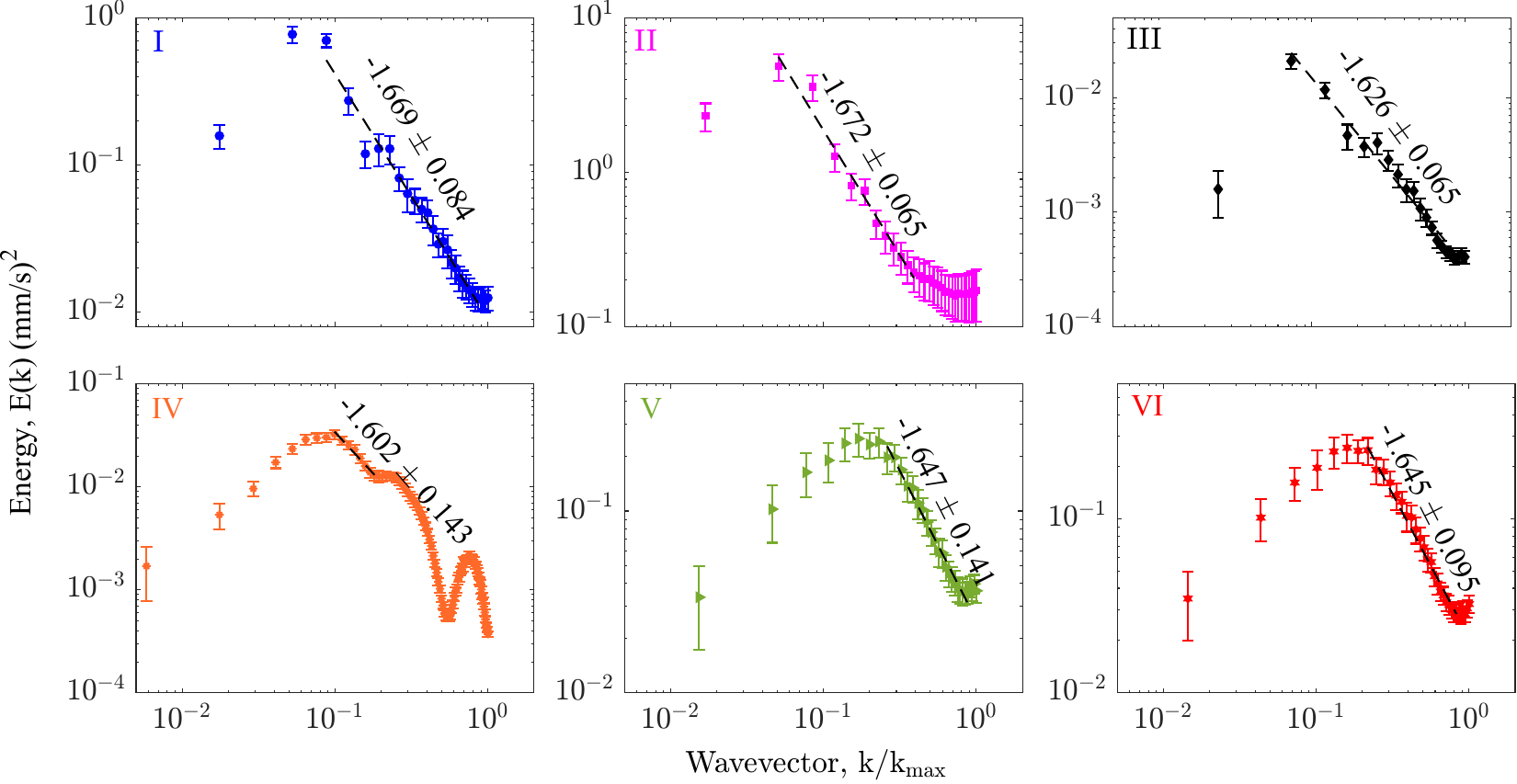}
\caption{The time-averaged energy spectra for Campaigns I–VI have been fitted with slopes that approximately match Kolmogorov's scaling for 3D turbulence. To present the spectra on a consistent scale, wavevectors are normalized by $\mathrm{k_{max}}$.
}
\label{Fig_3}
\end{figure*}
%%%%%%%%%%%%%%
%% font width is small, i will increase it%%
%%%%%%%%%%%%%%%%
The dust cloud in the diffused plasma region showed a variety of collective dynamics as we varied the discharge voltage between 230 and 600 volts. As listed in Table~\ref{table1_sets}, the self-sustained dust rotation has been observed for values at both edges of the voltage range. Figures~\ref{Fig_2} and \ref{Fig_222} show time-elapsed images for different campaigns where dust rotation is visible from particle trajectories. The inset in Fig.~\ref{Fig_2}(I) shows the tracking of selected particles in the box region. The particle tracking was performed using the ImageJ software's tracking algorithm. The exact location of each dust cloud with reference to the center of the cathode (in the Z-direction) and the surface of the glass (in the X-direction) is mentioned as axis labels. All other dust rotations are observed typically at the same location and similar discharge parameters except for campaigns I and II. For each campaign, the vortex motion is quite stable, and dust continues to rotate until we change the plasma creation conditions. We have provided dust rotation movies for each campaign as supplementary material~\cite{movies_suppli}. 

\paragraph*{}
We understand that a typical dust rotation in an unmagnetized dusty plasma may arise due to counteracting forces at spatially different locations~\citep{Bose_2019}. While gravity is always one force acting downwards, the radially outwards ion drag force in the void, the spatially varying force due to the sheath electric field are few possibilities in the counter direction, or at least their one component is opposite to the gravity. 
\paragraph*{}
A comprehensive analysis to establish the cause of dust rotation is beyond the scope of this paper. However, we ruled out some of the possibilities. A temperature gradient in neutral gas can give rise to thermophoretic force, which can affect dust dynamics \citep{PK_shukla_2015}. Typically, in glow discharge, the temperature of the cathode can increase reasonably due to ion bombardment after a few hours of operation \citep{Kaur_POP_2015a}. In our experiment, we observed the dust vortex soon after reaching the operating discharge voltage range before the cathode was reasonably heated. This suggests that thermophoretic force may not influence the vortex dynamics significantly. 
We also ruled out the possibility of a neutral flow in the vacuum chamber affecting the dust dynamics. We repeated the vortex experiment by simultaneously closing the pump and the Ar gas inlet to minimize any kind of neutral flow in the vacuum chamber. In this condition, too, we observed the dust vortex, thus ruling out any role of neutral flow.
\paragraph*{}
In earlier dusty plasma experiments with vortex observation, the causes involved variations in microparticle sizes~\citep{YOKOTA1996761,Rubin-Zuzic_2007}, charge gradients~\citep{Vaulina_2000,Vaulina_2003}, or a non-zero curl of the forces exerted by the plasma on the microparticles~\citep{Akdim_Goedheer_PRE_2003,Akdim_PRE_2003,Kaur_POP_2015b}. In the future, we plan to study the force balance comprehensively in our system. The goal of this paper is limited to demonstrating turbulence in a dusty plasma medium.
\paragraph*{}
Figures~\ref{Fig_4a} and \ref{Fig_4} provide flow velocity vectors and vorticity plots for all campaigns. The vorticity plots were obtained after taking the mean of velocity field profiles from 100 frames, which profoundly helped to visualize the large-scale rotations. 
The averaging cleans up the fluctuations and reflects the large-scale vortex structures.
%However, it washes out the information on small-scale vortex formations, 
However, these fluctuations are important when evaluating the energy spectrum in the next section.  
%In the previous paragraph, we anticipated different counteracting forces against gravity, causing the dust rotations. 
The vorticity profiles of all campaigns reflect that the dust rotation is clockwise for campaigns III-VI as shown in Fig.~\ref{Fig_4}, and the same is counterclockwise for campaigns I and II (see Fig.~\ref{Fig_4a}). This is observable from the velocity flow vectors and the colors of the vorticity profiles in different plots.

\subsection{The spatial energy spectra}
\label{spectrum_space}
%%%%%%%%%%%%%%%%%%%%%%%%%%%
%%%%%%%%%%%%%
\paragraph*{}
An important feature of turbulent flows is that they exhibit multi-scale energy cascades, which are typically demonstrated through the spatial energy spectrum \citep{kolmogorov_1991,frisch1995turbulence}.
Figure~\ref{Fig_3} presents subplots displaying the energy spectra for each campaign, which demonstrate the distribution of kinetic energy across different scales or wavenumbers.
The spectra are calculated in the following way:
For each recorded dataset, we take the PIV and get multiple velocity-flow profiles for each campaign.
Once we have a coarse-grained flow velocity ${\bf u(x},t)$ for a given snapshot, we perform a Fourier transform on ${\bf u(x},t)$ to obtain ${\bf u(k},t)$, from which we calculate the modal energy $E({\bf k},t) = |{\bf u(k},t)|^2/2$. After that, we find the one-dimensional shell spectrum $E(k,t)$ by adding up the modal energies within a certain range $k-1 \le k' \le k$, as described by~\citet{frisch1995turbulence}.
%%%%%%%%%%%%%%%%%%%%%%
\begin{figure*}
\centering
\includegraphics[width=0.9\textwidth]{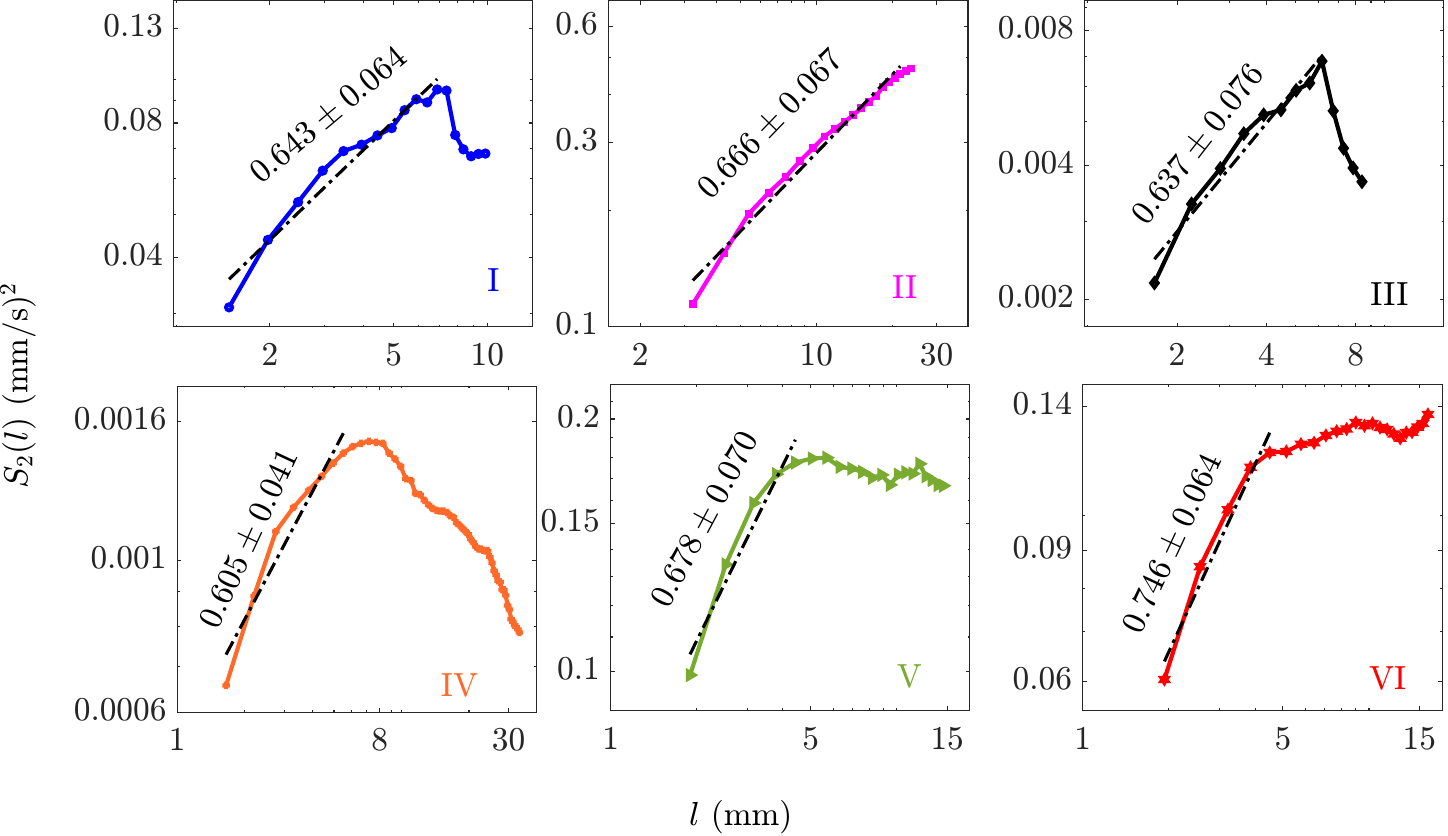}
\caption{Log-log plots of time-averaged second-order structure function $S_2(l)$ vs. incremental length $l$ for campaigns (I-VI). Each plot displays the results of different experimental setups, demonstrating variations in inertial range scaling and Kolmogorov scaling degree. The least squares fitted lines (black dashed) yield a slope with standard deviation.
}
\label{structure_function}
\end{figure*}
%%%%%%%%%%%%%
\paragraph*{}
We estimated the energy spectra for each flow profile of the recorded datasets separately. Then, we plotted the average energy spectra over several time frames with error bars, as seen in Fig~\ref{Fig_3} for all the campaigns. The wavevectors are normalized with \(\mathrm{k_{max}}\), where \(\mathrm{k_{max}}\) is the maximum wavenumber associated with the system. The values of \(\mathrm{k_{max}}\) for each campaign are \(6.334\), \(5.69\), \(5.69\), \(5.69\), \(4.91\), and \(4.91\) $\mathrm{mm^{-1}}$, respectively. Note that the units of $\mathrm{k_{max}}$ are $\mathrm{mm^{-1}}$, and the units of the energy spectra are \(\mathrm{(mm/s)^2}\). We perform normalization to present the spectra on the same scale without losing generality.
\paragraph*{}
We observe that the energy fed at the large scales cascades to the smaller scales following the standard -5/3 scaling for all campaigns, which is consistent with Kolmogorov's theory~\citep{kolmogorov1991dissipation}. The corresponding scaling in the real space is observed by analyzing the second-order structure function discussed in section~\ref{second_structure}.
However, note that the wavevector ranges vary slightly for different campaigns. Also, the total energy range in the spectra varies up to an order of magnitude as we look for the spectra for I-VI. It is based on the vortex domain size and finite flow velocities. For example, in Fig.~\ref{Fig_4a}, the vortex size of I is smaller than that of II; hence, its energy spectrum has a larger magnitude range.
\paragraph*{}
The flow dynamics is three-dimensional, but the present analysis is carried out only in the x-z plane due to the limitation of recording only one 2D plane using a laser sheet and a camera. However, we have individual movies of the y-z and x-z planes (asynchronous) for campaigns V and VI. Here, we provide in Fig.~\ref{PS_YZ_V} (see the Appendix), the energy spectrum and the vorticity plots for campaign V in the y-z plane. These plots demonstrate the 3-D characteristics of turbulent vortex flow in the cloud at the specific position of \(x = 5\) mm. The least-square fitted scaling (the slope) obtained is $-1.595 \pm 0.099$, consistent with our claims of Kolmogorov scaling within experimental uncertainties.
%%%%%%%%%%%%%%%%%%%%%%%%%%%%%%%%%%%%%%%%%%%%%%%%%
\subsection{Second-order structure factor}
\label{second_structure}
%%%%%%%%%%%%%%%%%%%%%%%%%%%%%%%%%%%%%%%%%%%%%%%%%%%%%%%%%
We analyze the structure-function for examining the statistical characteristics of turbulent flows \citep{frisch1995turbulence}. The structure functions display the moments of velocity increase. $\delta u^p_r = \langle[u(\vec r +l)-u(\vec r)] ^p\rangle$, where $r$ is a reference point, $p$ is the order of the structure function, and $l$ is the distance between two points in a velocity field~\cite{Chatterjee_SF-py_2023}. Based on Kolmogorov's theory (Kolmogorov 1941b), for fully developed homogeneous turbulence, the scaling relationships in the inertial range of turbulence should follow the form $\delta u^p_r \propto r^{\xi(p)}$, where $\xi (p) = p/3$ roughly;\citep{Benzi_1993_PRE}.
%%%%%%%%%%%%%%%%%%%%%%
%%%%%%%
\begin{figure*}[t]
\includegraphics[width =0.9\textwidth]{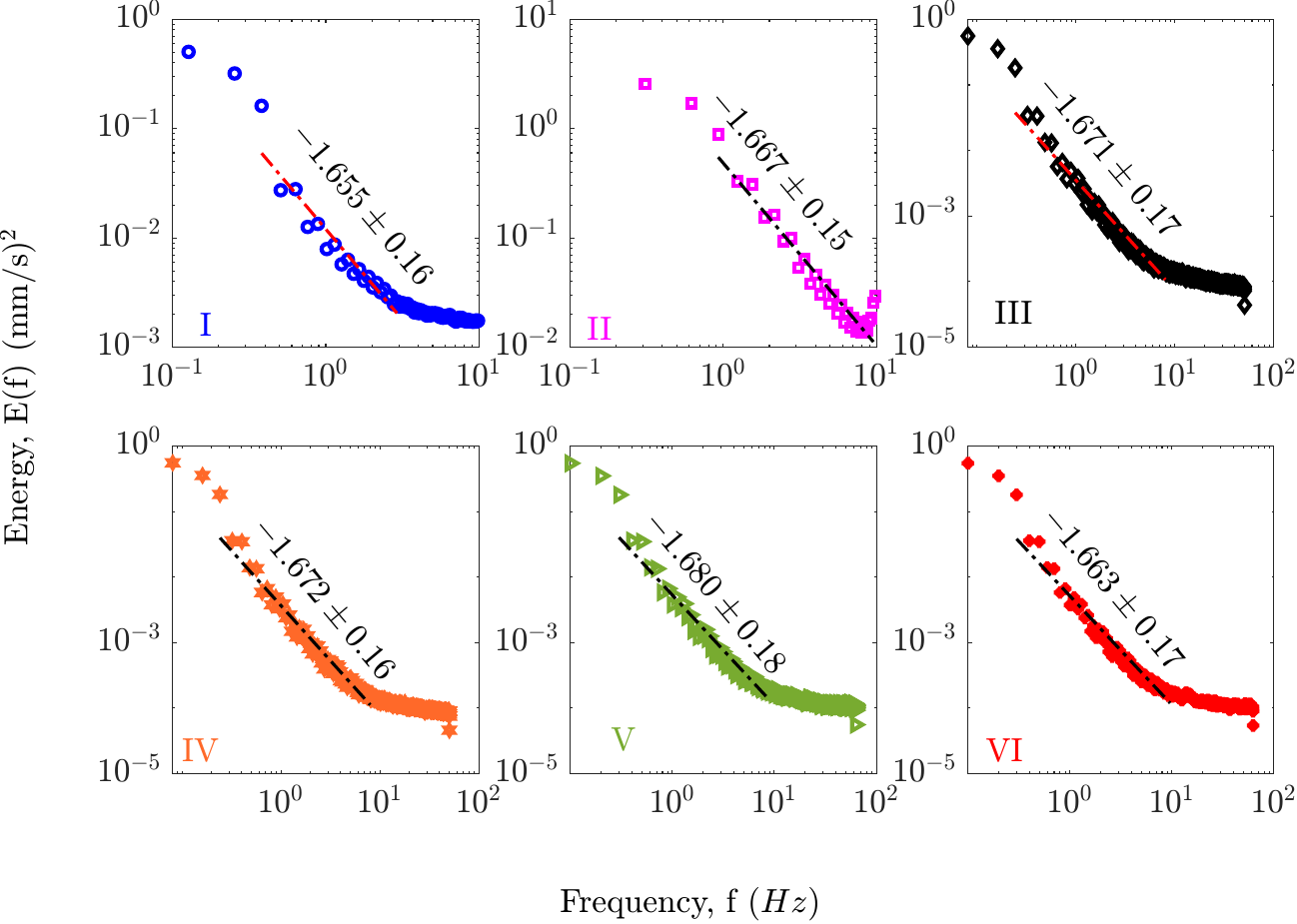}
\caption{Log-log plots of the frequency spectra are computed by taking the FFT to the time correlation of the velocity field. The spectra show a slight deviation from the Kolmogorov $-5/3$ prediction. 
% The error estimates for each spectrum are $0.46\%, 0.64\%, 3.34\%, 1.06\%, 3.1\%$, and $0.86\%$, respectively.
}
\label{frequency_spectra}
\end{figure*}
%%%%%%%%%%%%%%%%%%%%%%%%%%%%%%%%
\paragraph*{}
In this study, we calculate the second-order structure function for all data sets by determining the difference in velocity between a reference point $r$ and varying separation distances $l$, as shown in Fig.~\ref{structure_function}. To maintain generality, we computed the structural functions using a time average of 90 velocity frames. 
The fitted slopes yield the scaling exponent close to $0.66$ with standard deviation, which is almost consistent with Kolmogorov theory~\citep{kraichnan1974kolmogorov}.  We also see that the $S_2(l)$ values for campaigns III and IV are two orders of magnitude lower. This is because the velocity order is lower in these vortex regimes, which is also shown in the energy spectra, Fig~\ref{Fig_3}.

%%%%%
\subsection{The energy spectra of time-series}
\label{ps_time}
%%%%%%%%%%%%%%%%%%%%%%%%%
We also analyzed the energy spectra in the frequency domain, which gives energy distribution across various frequencies (wavelengths). This is done by taking the Fourier transform of velocity autocorrelation in accordance with the Wiener-Khinchan theorem ~\citep{frisch1995turbulence}.
\begin{equation}
E(f) = \frac{1}{2\pi} \int_{-\infty}^{\infty} e^{\iota f t} \langle v(t).v(t+\tau)\rangle~dt
\end{equation}
 We collect multiple velocity time series from different spatial locations, which usually fall on the dust rotation region. Subsequently, we followed the above methodology to compute the energy spectra for each time series. The mean frequency spectrum and the standard deviation have been provided for each campaign. We observed a Kolmogorov $-5/3$ power law over an inertial range of frequencies, as demonstrated in figure~\ref{frequency_spectra}. We computed the spectra for all six campaigns and found consistent scaling across all cases.
\paragraph*{}
Using analytical arguments, we also estimated the Kolmogorov dissipation length and time scales. These scales were calculated based on the mean energy transfer rate \(\epsilon \sim v_l^3/l\), where \(v_l = 2~\text{mm/s}\) is the average velocity and \(l = a\) is the average interparticle separation. The local viscosity is \(\nu \sim v_l l\). With \(\epsilon\) and \(\nu\), we computed the Kolmogorov length scale as \(l_d = (\nu^3/\epsilon)^{1/4} = 110~\mu \text{m}\) and the velocity scale as \(v_d = (\nu \epsilon)^{1/4} = 2~\text{mm/s}\). Consequently, the dissipative time scale is \(\tau_d = l_d / v_d = 0.0478~\text{s}\), leading to a dissipation frequency of approximately \(21~\text{Hz}\). This is also clear from Fig.~\ref{frequency_spectra}, which shows the dissipation frequency at almost $21~\mathrm{Hz}$.
%%%%%%%%%%%%%%%%%%%%%%%%%%%%%%%%
\begin{figure*}
\includegraphics[width = \textwidth]{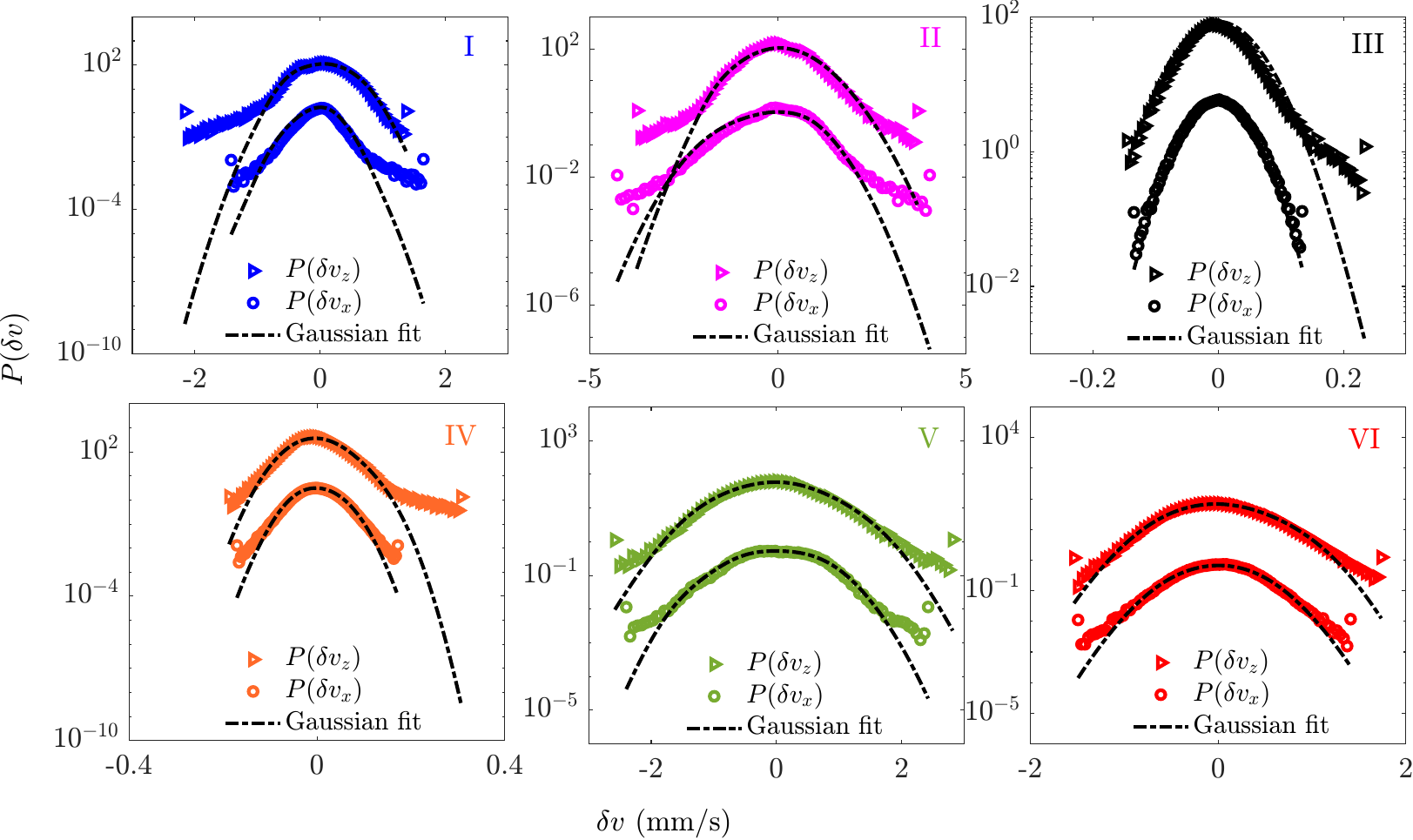}
\caption{
The PDFs of velocity gradients \(\delta v_x\) and \(\delta v_z\) for campaigns I-VI. The black dashed lines represent the corresponding Gaussian distribution fits. For improved visualization, we have artificially shifted the PDFs of \(\delta v_x\) by a factor of 0.01.
}
\label{pdf}
\end{figure*}
%%%%%%%%%%%%%
%
\subsection{Probability Distribution Functions}
\label{pdf_dust}
\paragraph*{}
Since the turbulent velocity fields exhibit significant fluctuations, it's essential to comprehend their statistical properties. Consequently, researchers have focused on studying velocity fields' PDFs and their derivatives. Velocity PDFs provide insights into the distribution of flow velocities at different points in space~\citep{pandit2009statistical, Supratik_PRF_2019}. The PDF of velocity fields should follow a Gaussian distribution in fully developed turbulent flows, but their derivatives often deviate from the Gaussian distribution, especially at the tails~\citep{GULITSKI_2007}.
\paragraph*{}
Figure~\ref{pdf} presents the PDFs of velocity gradients, \(\delta v_z\) and \(\delta v_x\), which deviate from the Gaussian distribution (shown by the black dashed lines) at the tails, indicating intermittency. Note that the PDFs shown are averaged over 90 datasets for each campaign. 
The fitted plots are obtained by providing $\delta v_x$ and $\delta v_z$ dataset to the standard MATLAB function ``fitdist" for ``normal" distribution. The distribution returns a mean value and the associated standard deviation in each case. Table~S2 detailing mean velocity, standard deviation, skewness, and kurtosis is included as supplementary material.
%%%%
\section{conclusion}
\label{conclusion}
In this work, we demonstrated the presence of Kolmogorov turbulence in a three-dimensional dusty plasma. The dust cloud formed naturally as a result of plasma and device configuration forces. The energy spectrum scaling of -5/3 and the second-order structure factor with a slope of 2/3 are reproducible for different campaigns, regardless of the vortices' origin and location. We could also demonstrate the characteristics of stationary turbulence due to -5/3 scaling in both spatial and time-series spectra. The self-consistent nature of the vortex motion of charged dust clouds also points to constant energy pumping up at the largest scales and dissipating as a thermal motion of dust particles. Overall, the dusty plasma medium acts as a driven dissipative system.
% \textcolor{blue}{An important aspect of the present work is the demonstration of fully developed turbulence in a low Reynolds number $R_e$ regime, which is typically between $8$ and $50$ in our case (see sec.~\ref{appd}). This is because dusty plasmas exhibit viscoelastic behavior. Earlier works also suggest the possibility of fully developed turbulence at a high Reynolds number~\citep {Tsai_Chang_I_2014,Zhdanov_2015}. Examining turbulence through multi-scale vortex dynamics in a dusty plasma at such low Reynolds numbers is an important phenomenon. Despite attempts to study turbulence in dusty plasma experiments in these low Reynolds number regimes, we still lack a proper energy transfer mechanism.}
\\
An important aspect of the present work is the demonstration of fully developed turbulence. Traditional understanding suggests the possibility of fully developed turbulence at a high Reynolds number $R_e$~\citep{kolmogorov1991local}. In dusty plasmas, earlier studies reported wave turbulence at low $R_e$ \citep{Tsai_Chang_I_2014, Zhdanov_2015}. 
The present work examines and reports fully developed turbulence through multi-scale vortex dynamics. 
Our calculations suggest typical values of $R_e$ between $8$ and $50$. The Section V appendix includes the calculation of the $R_e$. These results suggest dusty plasmas' similarity to viscoelastic fluids where elastic turbulence is observed at small $R_e$ values. This anticipation contradicts presented scalings from energy spectra and structure functions that suggest close alignment with Kolmogorov's prediction. This remains an open question, and we will try to examine this in future studies. 
\section*{Supplementary Material}
The following information is provided as part of the supplementary material:
(I) We provide an agreeable picture of full three-dimensional vortex flow using three different x-z planes at locations $y=-1,0,1$ mm. The energy spectra from all three planes are in the Fig.~S1. (II) The movies for campaigns I-VI have been provided in `.m4v' format. (III) The statistical parameters associated with the velocity gradients for campaigns I-VI.
\section*{Acknowledgements}
The authors acknowledge Prof. Mahendra Verma for his valuable discussions. ST acknowledges support from IIT Jammu PRAISE grant PRA-100003 for the experimental setup. ST and RW acknowledge partial support for this work through SERB Grant No. CRG/2020/003653. SB was supported by the U.S. Department of Energy under Contract No. DE-AC02-09CH1146. 
%%%%%%%%%%%%%%%%%%%%%%%%%%%%%%%%%%%%%%%%%%%%%%%%%%%%%%%%%%
% \bibliography{References_new,Dusty_plasma_ref,turb_rauoof,Turbulence_ref}
%%%%%%%%%%%%%
%merlin.mbs apsrev4-1.bst 2010-07-25 4.21a (PWD, AO, DPC) hacked
%Control: key (0)
%Control: author (8) initials jnrlst
%Control: editor formatted (1) identically to author
%Control: production of article title (-1) disabled
%Control: page (0) single
%Control: year (1) truncated
%Control: production of eprint (0) enabled
\bibliographystyle{aip}

%%%%%%%%%%%%%%%%%%%%%
{\section{Appendix}
\label{appd}
\subsection{Calculation for Reynolds number, $R_e$}
%%%%%%%%%%%%%%
\begin{figure*}
\centering
\includegraphics[width =0.8\textwidth]{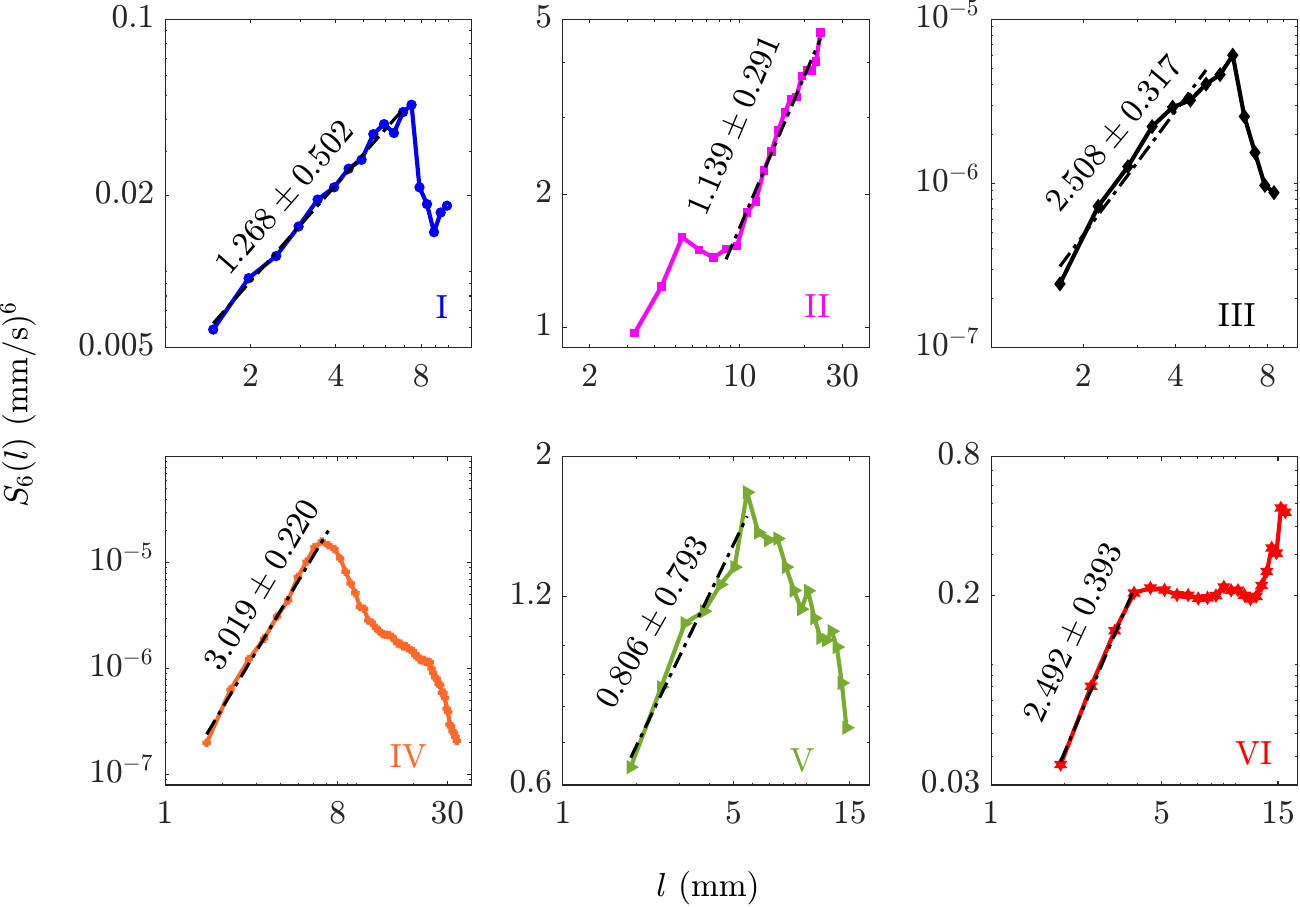}
\caption{Log-log plots of time-averaged sixth-order structure function $S_6(l)$ vs incremental length $l$ for campaigns I-VI. The least squares fitted lines (black dashed) yield a slope with standard deviation.}
\label{sixth_order_structure_function}
\end{figure*}
%%%%%%%%%%%%%%%%
\label{Reynolds_Number}
We calculated the Reynolds number, \(R_e\), for our experiments using the following parameters: \(\Gamma = 1500\), \(\kappa = 1.5\), \(m = 6 \times 10^{-13}\) kg, \(n = 2 \times 10^{10}\) m\(^{-3}\), \(\rho = 1.2 \times 10^{-2}\), \(a = 250 \mu\)m, and \(\omega_{pd} = 60\) Hz. From the literature, we took the normalized dynamic shear viscosity for 3D dusty plasma to be 0.2 ~\citep{Donko_PRE_2008,Hamaguchi_POP_2002,Murillo_PRL_2001}.

The normalization factor is \(\eta^* = mn\omega_{pd}a^2 = 4.5 \times 10^{-8}\) Pa.s.
Thus, the dynamic shear viscosity is \(\eta = 0.2 \times \eta^* = 9 \times 10^{-9}\) Pa.s.
Using a typical length scale \(l = 2 \times 10^{-2}\) m and a typical velocity scale \(v = 1 \times 10^{-3}\) m/s, we calculated the Reynolds number as $R_e = \rho lv/ \eta$ = 26. Depending on the range of parameter values, the Reynolds number in these experiments varies between 8 and 50.

\subsection{Measurement of intermittency, $S_6(l)$}
Because of intermittency~\citep{AGHA_1984}, the Kolmogorov $p/3$ scale in structure functions deviates at higher orders. In particular, the intermittency in turbulent flows is captured by the sixth-order structure-function. According to \citep{Vindel_NPG_2008, AGHA_1984}, the scaling exponent for the sixth-order structure functions differs from Kolmogorov's theory and is $(p/3 - \mu)$.
The sixth-order structure functions are plotted in our work, as Fig.~\ref{sixth_order_structure_function} illustrates. The intermittency parameters for campaigns I, II, and V are determined to be $\mu = 0.73$, $0.86$, and $1.2$, correspondingly. Nevertheless, the parameters for campaigns III, IV, and VI are $-0.5$, $-1$, and $-0.49$; for these campaigns, we need to perform more research as we do not yet have a satisfactory explanation.

%%%%%%%%%%%%%%%%%%%%%%%%%%%%%%
\subsection{The spatial energy spectrum in y-z plane}
%%%%%%%%%%%%%%%
We also provide the energy spectrum plot for campaign V in the Y-Z plane.
The Kolmogorov scaling is approximately visible in the intermediate range of wavenumbers (see Fig.~\ref{PS_YZ_V}). Note that the approximate location of the dust cloud is at x=5mm from the Y-Z plane.  
\begin{figure*}
\includegraphics[width = 0.8\textwidth]{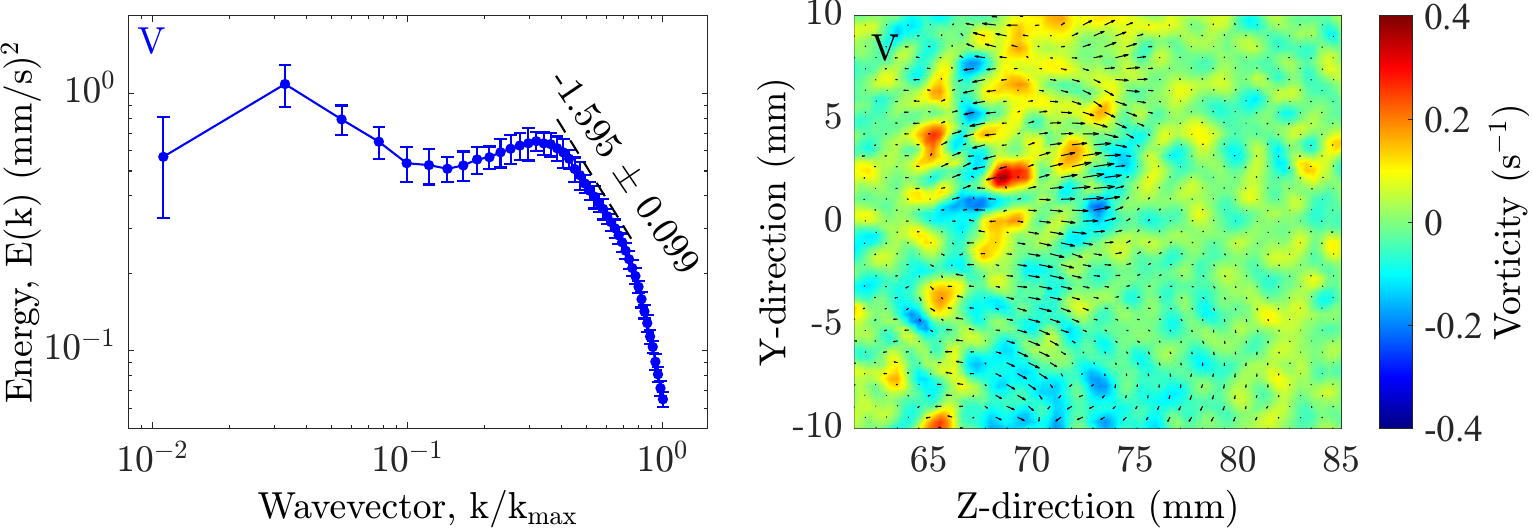}
\caption{Energy spectra and vorticity profiles in the Y-Z plane from campaign V. The approximate location in the X-direction is at x = 5 mm.}
\label{PS_YZ_V}
\end{figure*}
%%%%%%%%%%%%%
%%%%%%%%%%%%%%%%
\end{document}